\documentclass[12pt]{article}
\usepackage{epsfig}

\def\be{\begin{equation}}
\def\ee{\end{equation}}
\def\bea{\begin{eqnarray}}
\def\eea{\end{eqnarray}}
\def\lag{{\cal{L}}}

\def\DA{\partial_y \partial^y}
\def\dely{\partial_y}
\def\pe2{p_E^2}
\def\NR{\frac{n^2}{R^2}}
\def\NRB{\frac{\alpha^2}{R^2}}

\def\NRA{\frac{\alpha n}{R^2}}
\def\half{\frac{1}{2}}

\begin{document}
\newcommand{\mpl}{M_{\mathrm{Pl}}}
\setlength{\baselineskip}{18pt}
\begin{titlepage}
\begin{flushright}
KOBE-TH-01-07 \\
\end{flushright}

\vspace{1.0cm}

\centerline{{\LARGE\bf The Hosotani Mechanism in Bulk Gauge Theories }}  
\centerline{{\LARGE\bf with an Orbifold Extra Space S$^1$/Z$_2$ }} 
\vspace{25mm}

\centerline{Masahiro Kubo$^{(a)}$
\footnote{e-mail : kubo@phys.sci.kobe-u.ac.jp},
C. S. Lim$^{(b)}$
\footnote{e-mail : lim@phys.sci.kobe-u.ac.jp}
and Hiroyuki Yamashita$^{(a)}$ \footnote{e-mail : 
hy257@phys.sci.kobe-ac.jp}}
\vspace{1cm}
\centerline{$^{(a)}${\it Graduate School of Science and Technology, 
Kobe University,}}
\centerline{\it Rokkodai, Nada, Kobe 657-8501, Japan}
\centerline{$^{(b)}${\it Department of Physics, Kobe University,
Rokkodai, Nada, Kobe 657-8501, Japan}}
%
%
%
%
\vspace{2cm}
\centerline{\large\bf Abstract}
\vspace{0.5cm}

We pursue the possibility of the scenario in which the Higgs field is
identified with the extra-space component of a bulk gauge field.  The space-time we take is  M$^{4}$ $\otimes$ S$^1$/Z$_2$. 
We show that a non-trivial Z$_2$-parity assignment allows some of the 
extra-space component to have radiatively induced VEV, which strongly modifies 
the mass spectrum and gauge symmetry of the theory, realized by oribifolding. In particular we investigate the dynamical mass generation of zero-mode fermion and spontaneous gauge symmetry breaking due to the VEV. 
The gauge theories we adopt are a prototype model SU(2) and SU(3) model, of special interest as the realistic minimal scheme to incorporate the standard model 
SU(2) $\times$ U(1).  

\end{titlepage}

%
%

\section{Introduction}
One of the long standing issues in the particle physics is the hierarchy problem. Conventional wisdom to solve the problem in four dimensional scheme is to rely on the supersymmetry. Recently it has been realized that alternative solutions are possible once we extend our space time {\cite{Arkani, RS1, HIL}}. The authors of {\cite{Arkani, RS1}} adopted higher dimensional gravity theories, and aimed to solve the hierarchy problem between $M_{pl}$ and $M_{W}$, invoking to large extra dimension {\cite{Arkani}}, or to the ``warp factor" appearing in the non-factorizable AdS$_5$ space-time with 3-branes{\cite{RS1}}, though the hierarchy 
was discussed at the classical level. The approach taken in {\cite{HIL}} is a bit different: it deals with higher dimensional (bulk) gauge theories where the Higgs field is identified with the extra-space component of gauge field, and the main concern was the hierarchy problem at the quantum level. It has been shown that the summation of all Kaluza-Klein modes, inevitable to preserve the higher dimensional gauge symmetry, provides finite calculable Higgs mass, without suffering from the quadratic divergence  {\cite{HIL}}. See also {\cite{Daemi, Georgi,  Nomura, ABQ}} for recent related development in various contexts, utilizing the  finite Higgs mass due to the K-K mode sum.   

In this paper we pursue the possibility of the bulk gauge theory, with the Higgs field being identified with the extra-space component. This type of scenario has a remarkable feature that all interactions of the Higgs field are uniquely determined by the gauge principle. Thus we may hope that all arbitrariness coming from the presence of Higgs in the Standard Model may be removed. On the other hand, as the theory is so restrictive, to achieve desirable pattern of gauge symmetry breaking and observed mass spectrum is not easy. 
Another well-known serious problem concerning matter fields in bulk theories is that to get massless chiral fermions is difficult, in general. Namely, we easily get mirror fermions with opposite chirality and the same quantum number with known quarks and leptons, which easily form massive Dirac particles  with the ordinary matter.  

An interesting and powerful tool to solve these problems has been proposed 
{\cite{Kawamura, Kawamura2}} in the framework of  bulk gauge theory with orbifold S$^1$/Z$_2$ as its extra-space. (See {\cite{Rub-Sha, DS}} for other approaches based on localized fields on topological defects.) The key ingredient is that, in non-Abelian theories, we can assign non-trivial Z$_2$-parities for fields to form an irreducible repr. of the gauge group. For instance in D=5 SU(2) gauge theory, the theory is invariant under the Z$_2$ transformations of doublet fermion $\psi$, gauge fields $A_{\mu} (\mu = 0-3)$ and $A_{y}$ (extra space component, to be identified with Higgs): 
\be 
\psi \rightarrow P \gamma_{5} \psi, \ \ \ A_{\mu} \rightarrow P A_{\mu} P, \ \ \ A_{y} \rightarrow  -  P A_{y} P, 
\ee 
with $P^{2} = I$ ($I$: $2 \times 2$ unit matrix).  As an example let $P$ be diag$(1,-1)$.   
Then, only $\psi_{1R}$, $\psi_{2L}$ ($\psi_{1,2}$ are upper and lower components of the doublet), $A_{\mu}^{3}$ and $A_{y}^{1}, A_{y}^{2}$ ($1,2,3$: indices of  adjoint repr.) are  even functions of 5-th space coordinate $y$, and therefore have massless modes, when the theory is reduced to D=4 space-time. Thus as far as $1/R$ ($R$: the radius of $S^{1}$) is sufficiently large, the low energy effective 4-D theory has reduced gauge symmetry U(1) and massless chiral fermions (without mirror partners). In ref.{\cite{Kawamura}}, utilizing this mechanism in SU(5) GUT, the author has succeeded to solve so-called doublet-triplet splitting problem of 5-plet Higgs scalar, but at the classical level.  

In this paper we show that such realized gauge symmetry and mass spectrum by the method of ``Z$_2$ orbifolding"  illustrated above may be largely affected once the effect of the VEV of the extra-space component of gauge field $\langle A_{y} \rangle$, to be dynamically determined at quantum level, is included. Since the extra space S$^{1}$ has topologically non-trivial nature as a non-simply connected space, a constant background field $\langle A_{y} \rangle$ has non-trivial physical consequences, even though whose field strength is vanishing.    As the typical example, the mechanism to break gauge symmetry spontaneously by a Wilson loop $W = P \ \mbox{exp} (ig \oint \langle A_{y} \rangle dy)$ has been known as Hosotani mechanism {\cite{Hosotani}}. We show that the VEV $\langle A_{y} \rangle$, allowed in non-Abelian theories with non-trivial Z$_{2}$-parity assignment matrix $P$, generally causes not only spontaneous gauge symmetry breaking, but also the breaking of chiral symmetry, thus leading to dynamical mass generation for zero-mode fermions. At first sight the chiral symmetry seems never to be broken: the Z$_2$ symmetry acts as a chiral transformation, being accompanied by $\gamma_{5}$, on ferimon fields and this discrete symmetry seems to prevent fermions from developing any masses. 
We, however, should note that by a suitable choice of a non-trivial matrix P, we can arrange so that some  pair of right- and left-handed Weyl fermions has identical Z$_2$-parity 
(the example is the above-mentioned $\psi_{1R}$ and $\psi_{2L}$), and $A_{y}$, which is associated with the ``broken" generators connecting these Weyl fermions necessarily has even Z$_2$-parity. Thus the VEV $\langle A_{y} \rangle$ is allowed and generates dynamical fermion masses of the order $g \langle A_{y} \rangle$, without any contradiction with the Z$_2$ symmetry. 
Another interesting issue to note is that the $\langle A_{y} \rangle$, in spite of the fact that it is radiatively induced, is proportional to $1/(g R)$ and 
therefore generates dynamical masses of the order $1/R$, which are just comparable to the masses obtained at the classical level by the orbifolding. The reason  is rather easily understood as follows. The breaking of gauge and chiral symmetries are all controlled by the Wilson loop $W$. Just as in the case of the AB effect, when $W = I$, i.e. $g \langle A_{y} \rangle R = n \ (n: \mbox{integer})$, corresponding to the quantization condition of ``magnetic flux", the theory is equivalent to the case of $\langle A_{y} \rangle = 0$. Thus the effective potential as the function of the constant background $\langle A_{y} \rangle$ is a periodic function with a period $1/(g R)$. The expected $\langle A_{y} \rangle$ is therefore is expected to be of O$(1/(g R))$. 

If $1/R$ is much larger than the weak scale $M_{W}$, such quantum effect raises  the masses of some particles,  which we wish to remain as the members of the weak scale physics, to O$(1/R)$, and the low energy effective theory obtained by the orbifolding is largely affected. For instance,  the Higgs doublet, which remained massless after the ``doublet-triplet splitting" by orbifolding {\cite{Kawamura}}, may get a huge finite dynamical mass of O$(M_{GUT})$, which is undesirable from the view point of gauge hierarchy problem. (Note that in ref.{\cite{Kawamura}}, the Higgs scalar  is not regarded as an extra-space component of gauge field and the Higgs mass-squared@gets quadratically divergent quantum correction, anyway.) 
If, on the other hand, $1/R$ is comparable to $M_{W}$, this mechanism utilizing  $\langle A_{y} \rangle$ may potentially provide a useful way to realize spontaneous gauge symmetry breaking together with the mass generation for ordinary matter fields through Yukawa coupling, which are essential ingredients needed in the 4-D Standard Model. We discuss, in addition to a prototype model SU(2), SU(3) model as a realistic minimal scheme to incorporate the standard model SU(2) $\times$ U(1), where such nice features of $\langle A_{y} \rangle$ manifest themselves.

\section{Analysis at classical level and low energy effective theory} 
Our eventual goal should be to construct a realistic model of elementary particles, starting from a bulk gauge theory with matter fermions, which also propagate 
the bulk space-time, but without any higher dimensional scalars. The 4-dimensional (4D)  scalar particle, the Higgs, should be identified with the extra space component of the bulk gauge field. The extra space is assumed to be an orbifold S$^{1}$/Z$_2$, which plays key role for the construction. In this section the analysis is made at the classical level.
      
The theory we focus on is 5D SU(N) Yang-Mills theory with N$_f$ 
fermions belonging to the fundamental repr. of SU(N). The space-time to work with is M$^{4} \times$ S$^1$/Z$_2$. We first leave N to be arbitrary for generality. The lagrangian reads as  
\be
\lag={\displaystyle{\sum_{f=1}^{{\rm{N}}_f}}}
\bar{\psi^{(f)}}i\gamma^M D_M \psi^{(f)} -\half {\rm{tr}}(F_{MN}F^{MN})+
\lag_{gf}+\lag_{gh},
\ee
where we denote the gauge-fixing and ghost terms by $\lag_{gf}$ and 
$\lag_{gh}$. 5D  space-time coordinates, gamma matrices and covariant derivative are denoted as $x^M=(x^{\mu}, y)$, $\Gamma^M \equiv (\gamma^{\mu},i\gamma_5)$ ($\mu = 0-3$), and $D_M=\partial_{M} + igA_M$, respectively. \footnote{ We use the metric $g_{MN}=$ diag$(1,-1,-1,-1,-1)$ and the notation $A_M={\displaystyle{\sum_{a}}} A_M^a T^a$ and $F_{MN}={\displaystyle{\sum_{a}}} F_{MN}^a T^a$, where the generators of gauge group is normalized as ${\rm {tr}}(T^aT^b)=\half \delta_{ab}$.}  

As the extra-space is the orbifold S$^1$/Z$_2$, the lagrangian should be invariant under $y \rightarrow y+2 \pi R \ (R: \mbox{the radius of S}^{1})$ and $y \rightarrow -y$. Because the theory under consideration has a continuous global symmetry U(N), the invariance under $y \rightarrow y+2 \pi R$ does not necessitate periodic boundary conditions for fields and  it is generally possible to impose a ``twisted boundary conditions"  $\psi(x,y+2\pi R)=U\psi(x,y)$ and $A_M(x,y+2\pi R)=UA_M(x,y)U^{\dagger}$ where $U$ is a $y$-independent unitary matrix. 
\footnote{Strictly speaking, the theory also has a global symmetry U(${\rm{N}}_f$) among ${\rm{N}}_f$ flavors, orthogonal to U(N). However, we will not consider the twisting associated with U(${\rm{N}}_f$), since it cannot be reduced to any constant background of extra space component of gauge field and therefore is not dynamically determined.} Similarly the following ``twisted" identification under the Z$_2$-parity transformation is possible {\cite{Kawamura}}, 
\be
\psi^{(f)} (x,-y) = \gamma_5 \ P \psi^{(f)} (x,y); \ \ A_{\mu}(x,-y) = P A_{\mu}(x,y) P, \ \ A_{y}(x,-y) = - P A_{y}(x,y) P. 
\ee
The matrix $P$ is a constant unitary matrix with a condition $P^{2}= I \ (I$: N $\times$ N unit matrix), which also imply that $P$ is hermitian. 
\footnote{We again ignore the twisting associated with U(${\rm{N}}_f$).} 
Let us note that bare mass term $\bar{\psi^{(f)}} \psi^{(f)}$ is incompatible with the Z$_2$-parity symmetry, irrespectively of the choice of $P$.   

For the ``twisted" boundary condition to be consistent with the 
Z$_2$ symmetry, a consistency condition  
\be
U P U= P. 
\ee
should be imposed. Actually if we further utilizes the freedom of a local gauge transformation with gauge parameter, linear in $y$ , the twisted boundary condition can be removed except for a U(1) factor. Namely, writing $U$ as $U = e^{i \phi} \mbox{exp} (i \theta^{a}T^{a})$, a gauge transformation, 
\bea 
\psi^{(f)} \rightarrow \psi^{(f) \prime} &=& \mbox{exp}(- i \frac{\theta^{a}T^{a}}{2\pi R} y) \ \psi^{(f)} ,  \nonumber \\ 
A_{M} \rightarrow A^{\prime}_{M} &=& \mbox{exp}(- i \frac{\theta^{a}T^{a}}{2\pi R} y) \ A_{M} \ \mbox{exp}( i \frac{\theta^{a}T^{a}}{2\pi R} y) + \delta^{y}_{M} \frac{\theta^{a}}{2\pi gR} \cdot T^{a},
\eea 
makes the fields $\psi^{(f) \prime}, A^{\prime}_{M}$ satisfy periodic boundary condition, up to an overall U(1) phase factor, i.e. $\psi^{(f)\prime} (x, y+ 2\pi R) = e^{i \phi} \psi^{(f) \prime} (x, y)$. The remaining phase factor, $U = e^{i \phi} \cdot I$, however, is incompatible with the consistency condition (4), 
unless $e^{i\phi} = \pm 1$.  Thus we can assume $U = I$, periodic boundary condition,  without loss of generality \footnote{Another possibility $U = - I$ is not of our interest, as it forbids the presence of fermion zero mode.}, though  such ``large gauge transformation" should be compensated by a constant shift of the background   field, $A_{y}^{a} \rightarrow A_{y}^{a} + \theta^{a}/(2\pi gR)$. Note that since S$^{1}$ is non-simply connected space, such constant background is physically meaningful and should be determined dynamically at quantum level, as we will see below. Further using a remaining freedom of global U(N) transformation the matrix $P$ always can be diagonalized .(Recall that $P$ is hermitian.) Thus without loss of generality we can set $P =$ diag$(1,...,1,-1,...,-1)$. 

 In the rest of this section, we will study at the classical level how the choice of Z$_2$-parity matrix $P$ strongly affects  the mass spectrum of bulk fields, thus determining the structure of the low energy effective theory.

\subsection{SU(N) gauge theory with trivial Z$_2$-parity assignment}

The first example is the case of trivial Z$_2$-parity matrix, $P = I$. Each field therefore has the following Z$_2$-parity:  
\be
A_{\mu}^{a}(x,-y)= A_{\mu}^{a}(x,y), \ \ 
A_{y}^{a}(x,-y)= -A_{y}^{a}(x,y), \ \ 
\psi_j(x,-y)= \gamma_5\psi_j(x,y)
\ee
where $a$=1,2,..,{\rm{N}}$^{2}$-1, and $j = 1$ to $N$. Flavor index $f$ is suppressed for a while till it becomes necessary.  To be consistent with these parity assignment, the gauge parameter $\omega^{a}$ should behave as 
\be
\omega^{a}(x,-y)= \omega^{a}(x,y) (=\omega^{a}(x,y+2\pi R)).
\ee
Thus for all indices $a$ four-dimensional gauge transformation with gauge parameter, independent of $y$, is allowed, and full gauge symmetry SU(N) remains in the 4D low energy effective theory. In fact the above parity assignment allows all $A_{\mu}^{a}$ to have zero(y-independent)-modes $A_{\mu}^{a(0)}(x)$ and they all remain massless in the reduced 4 D space-time. On the other hand we learn that the  zero-modes of $A_y^{a}$ are absent. 
Thus if excited K-K modes with masses $n/R \ (n \neq 0)$ are regarded to decouple from the low energy world, the 4D low energy effective theory has no scalar field. 

Concerning fermions, because of the presence of $\gamma_{5}$ in the Z$_{2}$ 
transformation, we know only the right-handed components have zero-modes, irrespectively of $j$. More explicitly, in the 2-component notation of Weyl fermion, each $\psi_{j}$ can be expanded in Fourier series as  
\be
\psi_j(x,y)=\frac{1}{\sqrt{2\pi R}}\pmatrix{u^{(0)}_{j,R}(x)+
{\displaystyle{\sum_{n=1}^{\infty}}} u^{(n)}_{j,R}(x) \sqrt{2}
\cos \frac{ny}{R} \cr
{\displaystyle{\sum_{n=1}^{\infty}}} u^{(n)}_{j,L}(x) \sqrt{2}
\sin \frac{ny}{R}
}. 
\ee
Thus the reduced 4D low energy effective theory is SU(N) gauge theory containing a full multiplet of gauge field $A_{\mu}^{(0)} (x)$ and a full multiplet of massless Weyl fermion of the same chirality $\psi^{(0)}_{R} (x)$, but without any scalar (Higgs) field.

\subsection{SU(N) gauge theory with non-trivial Z$_2$-parity assignment} 

Next we discuss the case of non-trivial Z$_2$-parity assignment, with $P$ not being proportional to $I$; $ P =$ diag$(1,..,1,-1,..,-1)$ with $N_{+}$ elements of 1 and $N_{-}$ elements of -1 ($N_{+} + N_{-} = N$; \ $N_{+}, N_{-} \geq 1$). 
If $[T^{a}, P] = 0$ , $P T^{a} P = T^{a}$ and the associated gauge fields $A_{\mu}^{a}$ have  zero-modes. Thus this choice of $P$ breaks the gauge symmetry as  

G = SU(N) $\rightarrow$ H = SU($N_{+}$) $\otimes$ SU($N_{-}$) $\otimes$ U(1). \\
Note that the rank of the gauge group is not reduced. For a special case of $N_{+}, N_{-} = 1$, SU(1) should be just ignored.  

Accordingly, zero-modes, remaining in low energy 4D effective theory is the following: 
\be 
A_{\mu}^{a(0)} \ (a \in H); \ \ 
A_{y}^{a(0)}  \ (a \in G/H); \ \ 
\psi_{jR}^{(0)} \ (j= 1 \ \mbox{to} \ N_{+}), \ \ 
\psi_{jL}^{(0)} \ (j= N_{+}+1 \ \mbox{to} \ N). 
\ee   
We also note that zero-modes of gauge parameters $\omega^{a}$ appear only for $a \in H$, to be consistent with the remaining gauge symmetry. 

In contrast to the case of $P = I$, now fermion zero-modes have both chiralities, and once $A_{y}^{a(0)} \ (a \in G/H)$, connecting Weyl fermions with different chiralities, develop non-vanishing VEV, these Weyl fermions will form a massive Dirac particle, as we will see in the following section for the details. The dynamical masses are produced  through the 5D gauge interaction (4D Yukawa couplng),      
\be 
ig \bar{\psi}\gamma_5 \langle A^{a(0)}_{y}\rangle T^{a} \psi \ \ \ (a \in G/H). \ee

\section{One-loop effective potential and the Hosotani mechanism in bulk gauge theories with non-trivial Z$_2$-parity assignment}

In this section we discuss how the dynamical fermion mass generation and spontaneous gauge symmetry breaking is realized via the radiatively induced VEV of $A_{y}$, i.e. the Hosotani mechanism {\cite{Hosotani, Hatanaka}}. As has been already pointed out, in bulk gauge theories with extra orbifold space, such mechanism is possible only for the case with non-trivial Z$_2$-parity assignment. 

The bulk gauge theories we consider here are SU(2) gauge theory, as a prototype model, and SU(3) model, which is quite interesting in the sense that it provides realistic minimal framework to incorporate the standard model SU(2) $\times$ U(1).  

The  VEV of $A_{y}$ is a finite calculable quantity, which is determined by the  minimization of the one-loop induced effective potential as the function of constant background field $\langle A_{y}\rangle \equiv B_{y}$; we utilize background field method.

It is convenient to work with a gauge-fixing term for quantum fluctuation $\tilde{A}_{M}$, which uses gauge covariant derivative concerning the background field $B_{M}$, $\frac{1}{\xi} \mbox{tr} (D_{M} \tilde{A}^{M})^{2}, \ D_{M} \tilde{A}^{M}= \partial_{M} \tilde{A}^M+ig[B_M, \tilde{A}^M]$ in the Feynman-'t Hooft gauge, $\xi = 1$.  

\subsection{SU(2) model}
We first take the SU(2) bulk gauge theory with $P = $diag $(1,-1)$, as the prototype model with non-trivial Z$_2$-parity assignment.     
At the classical level, the gauge symmetry is broken by the orbifolding: 
\be 
G = \mbox{SU(2)} \rightarrow H = \mbox{U(1)},  
\ee 
where the remaining U(1) is due to the generator $T^{3}$.  
The zero-modes, remaining in low energy 4D effective theory is the following: 
\be
A_{\mu}^{3(0)};  \ \ 
A_{y}^{1,2(0)};  \ \ 
\psi_{1R}^{(0)}, \ \ 
\psi_{2L}^{(0)} . 
\ee   
Thus only $A_{y}^{1,2(0)}$ may develop VEV, i.e. $B_y^{a} = (B_y^1, B_y^2, 0)$. Thanks to the remaining global U(1) symmetry, we can always put this into the form of $B_y^a=(B_y^1, 0, 0)$ . 
The 4D mass-squared operator $D_y D^y$ for the fields in the adjoint repr. under this background is 
\be
[D_yD^y]_{ab}=\pmatrix{\DA & 0 & 0 \cr
         0 & \DA-g^2B_y^1B^{y1} & -2gB^{y1} \dely \cr
         0 & 2gB^{y1}\dely & \DA-g^2B_y^1B^{y1}
},
\ee
where $a, b = 1,2,3$ are adjoint indices.
Since this operator does not mix different K-K modes, we can diagonalize this operator separately for each K-K mode $n$. For $n \neq 0 \ (n > 0)$, the matrix elements of  the operator is easily obtained by using the basis of orthonormal functions $\frac{1}{\sqrt{\pi R}} \cos{(\frac{n}{R}y)}$ for $a,b = 3$, and $\frac{1}{\sqrt{\pi R}} \sin{(\frac{n}{R}y)}$ for $a,b =1,2$:  
\be 
\pmatrix{
          \NR & 0             & 0             \cr
          0       & \NR+\NRB & 2\NRA         \cr
          0       & 2\NRA         &    \NR+\NRB
},  
\ee 
where $\alpha \equiv g B^{1}_{y} R$. The eigenvalues are readily known to be $\NR, \ \frac{(n+\alpha)^{2}}{R^{2}}, 
\ \frac{(n-\alpha)^{2}}{R^{2}}$.  
For the zero-mode $n = 0$, only $A_{\mu}^{3(0)}$ with 4D mass-squared 
$\NRB$ exists. 
We note that 4D vector fields $A_{\mu}$ and F.-P. ghosts have the same Z$_2$-parity assignment, and the effect of the ghosts is just to reduce the degree of the polarization to the physical one, $4 \rightarrow 2$. Thus, combining the contributions from both sectors of zero and non-zero modes, we obtain the contribution of 4D vector and ghost fields,  
\be
V_{eff}^{v+g}=\frac{1}{2 \pi R} \int \frac{d^4 p_E}{(2 \pi)^4}
\half \cdot 2 [{\displaystyle{\sum_{n=1}^{\infty}}} \log(\pe2+\NR)
+{\displaystyle{\sum_{n=-\infty}^{\infty}}} \log(\pe2+(\frac{n-\alpha}{R})^2)], \ee
where $p_{E}$ is Euclidean momentum. 
In the case of 4D scalar $A_{y}$, the mass-squared matrix is the same as those of $A_{\mu}$ and F.-P. ghosts, except that the Z$_2$-parity assignment is just opposite. Thus for zero-modes $A_{y}^{1,2(0)}$, the mass-squared matrix is diag$(0, \NRB)$. The contribution of $A_{y}$ is thus given as 
\be
V_{eff}^{s}=\frac{1}{2 \pi R} \int \frac{d^4 p_E}{(2 \pi)^4}
\half \cdot 1 [{\displaystyle{\sum_{n=0}^{\infty}}} \log(\pe2+\NR)
+{\displaystyle{\sum_{n=-\infty}^{\infty}}} \log(\pe2+(\frac{n-\alpha}{R})^2)].
\ee
Summing up these two contributions, we obtain the one-loop induced effective potential from 4D vector, ghost and scalar fields, 
\bea
V_{eff}^{v+g+s}=&& \frac{1}{2 \pi R} \int \frac{d^4 p_E}{(2 \pi)^4}\nonumber\\
&& \times {\Large (}
\half \cdot 3 [{\displaystyle{\sum_{n=1}^{\infty}}} \log(\pe2+\NR)
+{\displaystyle{\sum_{n=-\infty}^{\infty}}}
\log(\pe2+(\frac{n-\alpha}{R})^2)] 
+\half \log \pe2 {\Large )}.
\eea
By a suitable reguralization method and subtracting irrelevant $\alpha$-independent part we obtain a finite result {\cite{Hosotani}}. 
\be
V_{eff}^{v+g+s}=-3C{\displaystyle{\sum_{n=1}^{\infty}}} \frac{1}{n^5}
\cos (2\pi n \alpha),
\ee
where $C=\frac{3}{128 \pi^7 R^5}$.

On the other hand, the mass-squared operator for fermions belonging to a fundamental repr. is 
\be
[D_y D^y]_{ij}=\pmatrix{
          \DA+\frac{\alpha^2}{4R^2} & -i\frac{\alpha}{R}\dely \cr
          -i\frac{\alpha}{R}\dely & \DA+\frac{\alpha^2}{4R^2} \cr
}.
\ee 
For non-zero K-K modes, the matrix in the basis of Dirac fermion $\psi_{1}, \ \psi_{2}$reads as 
\be
[D_y D^y]_{ij}=\pmatrix{
          \NR +\frac{\alpha^2}{4R^2} & -i\frac{\alpha n}{R^{2}}\gamma_{5} \cr
          i\frac{\alpha n}{R^{2}}\gamma_{5}  & \NR+\frac{\alpha^2}{4R^2} \cr
}, 
\ee 
whose eigenvalues are $\frac{(n+\frac{\alpha}{2})^{2}}{R^{2}}$ and $\frac{(n-\frac{\alpha}{2})^{2}}{R^{2}}$, each having 4-fold degeneracy. The zero-mode is composed of two Weyl fermions $\psi_{1R}$ and $\psi_{2L}$ and the mass-squared matrix in this basis reads as  
\be
[D_y D^y]_{ij}=\pmatrix{
          \frac{\alpha^2}{4R^2} &  0 \cr
          0  & \frac{\alpha^2}{4R^2}  \cr
}. 
\ee 
We thus find the contribution of N$_{f}$ fermions to the effective potential 
\be 
V_{eff}^f =  - \mbox{N}_{f} \cdot \frac{1}{2 \pi R} \int \frac{d^4 p_E}{(2 \pi)^4}
\half \cdot 4 [ {\displaystyle{\sum_{n=-\infty}^{\infty}}} 
\log(\pe2+(\frac{n-\alpha/2}{R})^2)].
\ee 
The relevant finite part is 
\be
V_{eff}^f=4{\rm{N}}_fC{\displaystyle{\sum_{n=1}^{\infty}}} \frac{1}{n^5}
\cos (\pi n \alpha).
\ee

\begin{figure}
\centerline{
\epsfxsize=6.5cm
\epsfbox{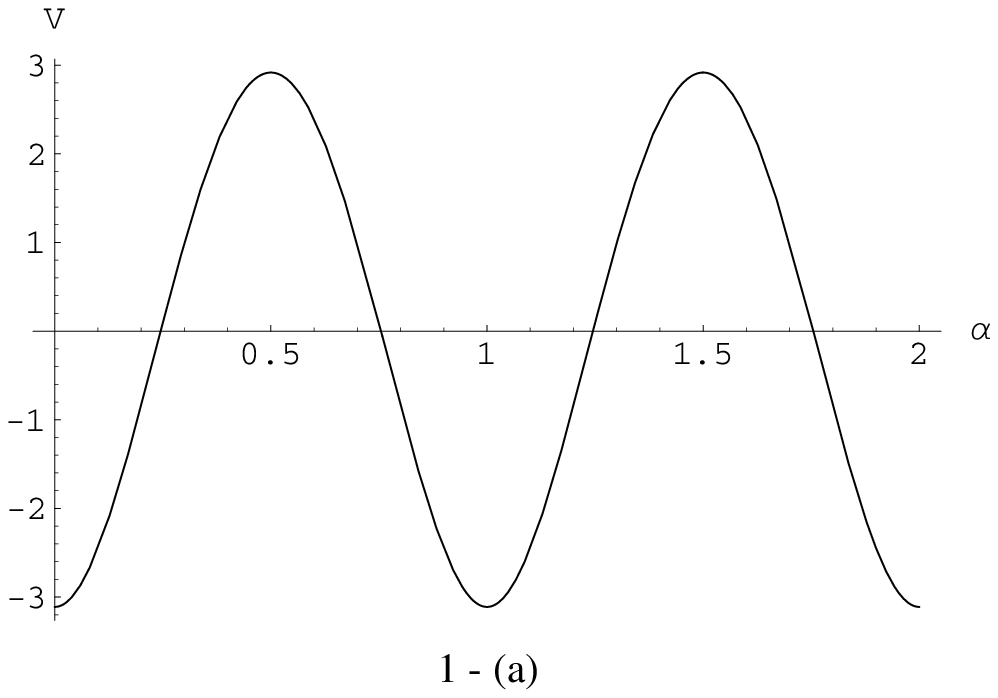}
\epsfxsize=6.5cm
\epsfbox{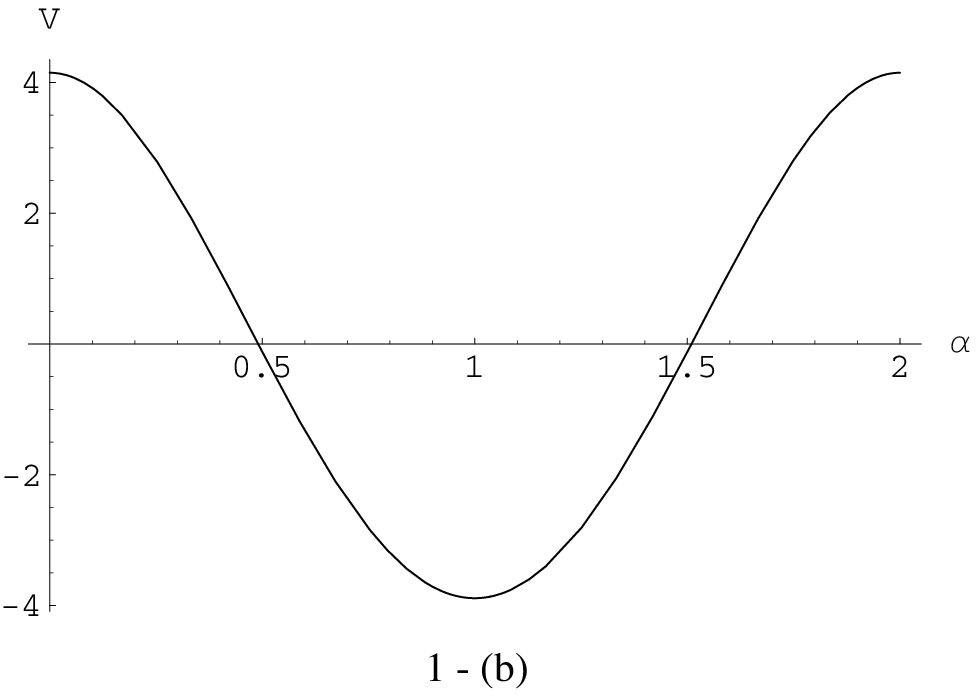}
}
\caption{The one-loop effective potential in the unit of $C=\frac{3}{128 \pi^7 R^5}$ as a function of $\alpha = g B^{1}_{y} R$ in the SU(2) model: (a) The contribution from gauge-ghost system, $V^{v+g+s}_{eff}$. (b) The contribution from fermions $V^{f}_{eff}$.}
\end{figure}

Combining all contributions, we get the effective potential  
\be
V_{eff} (\alpha) = V_{eff}^{v+g+s} + V_{eff}^f 
= C{\displaystyle{\sum_{n=1}^{\infty}}} \frac{1}{n^5} [-3 \cos (2\pi n \alpha) +4{\rm{N}}_f \cos (\pi n \alpha)]. 
\ee 
As we learn from Fig.1, though for pure Yang-Mills case ${\rm{N}}_f = 0$ the theory has  two degenerate vacua at $\alpha=0$ and $1$, including fermions the global minimum is located at $\alpha = 1$. We thus obtain VEV $B_{y}^{1} = 1/(gR)$, which  may spontaneously break gauge symmetry (note that $[ T^{3},B_{y}] \neq 0$) and simultaneously generates dynamical fermion mass for the zero mode. 
Concerning the spontaneous gauge symmetry breaking, however, we find that actually the U(1) symmetry is not broken by the VEV $B_{y}$. This is because what we should care about is the commutator of the Wilson-loop $W$. We note $\langle W \rangle = P \ \mbox{exp} (ig \oint B_{y} dy) = \mbox{exp} (ig B_{y}^{1} \frac{\sigma_{1}}{2}(2\pi R)) = \mbox{exp} (i \pi \sigma_{1}) = - I$. Thus $[ T^{3},\langle W \rangle] = 0$ and the gauge symmetry is not spontaneously broken. We can explicitly confirm this from the mass-squared matrix of $A_{\mu}$. Though the mass-squared of the zero mode $A_{\mu}^{3(0)}$ is raised to $\frac{1}{R^{2}}$ due to the VEV, we learn that the mass-squared matrix for $n = 1$ sector now has zero eigenvalue: 
$\frac{(1-1)^{2}}{R^{2}} = 0$ for $\alpha = 1$. 

On the other hand, concerning fermion masses, what matters is whether $\langle W \rangle = I$  or not. As $\langle W \rangle = - I$, we expect that fermions obtain a dymnamical Dirac mass. 
It is actually the case and we have a mass term for the zero mode (for every flavor index $f$)
\be 
ig \bar{\psi^{(0)}}i \gamma_5 B^{1}_{y} T^{1} \psi^{(0)} = 
-i \frac{g}{2} B^{1}_{y}[\bar{\psi_{1R}^{(0)}} \psi_{2L} - \bar{\psi_{2L}^{(0)}} \psi_{1R}]. 
\ee
Thus the generated dynamical mass is $\frac{g}{2} B^{1}_{y} = \frac{1}{2R}$. 
In fact the mass-squared matrix for the fermion zero-mode, discussed above,  has eigenvalue  $\frac{1}{4R^{2}}$ with $\alpha = 1$, while the matrix for $n =1$ mode also has an eigenvalue $\frac{(1 - \frac{1}{2})^{2}}{R^{2}}$, but no 0 eigenvalue. 

 The same mechanism is known to work for matter scalar fields as well, as long as they belong to a fundamental repr., $(\phi_{1}, \phi_{2})^{T}$. The same mass-squared matrix as the case of fermions (except for spin degree of freedom and $\gamma_{5}$ factor) is obtained for $n \neq 0$ modes, though in the case of scalar only $\phi_{1}$ is allowed to have a zero-mode, whose mass-squared eigenvalue is again dynamically raised to $\frac{1}{4R^{2}}$.  The same mechanism is expected to be operative even in the case of 5-plet Higgs fields of SU(5) GUT,  and the doublet-triplet splitting, realized at the classical level {\cite{Kawamura}}, is expected to be spoilt once quantum effect is switched on, as the doublet scalar (corresponding to $\phi_{1}^{0}$ in our model) is expected to get radiatively  generated  mass of the order $\frac{1}{R} \sim M_{GUT}$.

In addition, $A_{y}^{1}$ itself, which we identified with the Higgs, also obtain the finite mass {\cite{HIL}} 
\be 
m^2_{A_y} = [g^2R^2\frac{\partial^2 V_{eff}(\alpha)}
{{\partial \alpha}^2}]_{\alpha=1}
  = \frac{9 g^2}{128 \pi^5 R^3}(4+{\rm{N}}_f)\zeta(3) 
 = \frac{9 g_{4}^2}{64 \pi^4 R^2}(4+{\rm{N}}_f)\zeta(3),
\ee 
where $\zeta (z)$ is the Riemann's zeta-function and $g_{4} = g/(\sqrt{2\pi R})$ is the 4D gauge coupling.

\subsection{SU(3) model}
One may naively expect that a realistic model incorporating the standard model 
can be constructed just putting the gauge theory SU(2) $\otimes$ U(1) in the bulk. We, however, immediately notice it is not the case; 4D scalar $A_{y}$, to be identified with Higgs field, belongs to the adjoint repr. 3 of SU(2), but not to the doublet, which the Higgs should belong to. Such problem is evaded once we enlarge the gauge group a little bit. The possible minimal extension is the SU(3) bulk gauge model. Let us note that the adjoint of SU(3) decomposes into the repr.s of subgroup SU(2) as $8 \rightarrow 3 + 2 +2 +1$. 

The SU(3) symmetry can be broken by orbifolding into that of the standard model, once we adopt a non-trivial Z$_2$-parity assignment $P = \mbox{diag} (1, -1, -1)$: 
\be 
G = \mbox{SU(3)} \rightarrow H = \mbox{SU(2)} \otimes \mbox{U(1)}.  
\ee  
There is a bonus; among 3 + 2 + 2 + 1 of $A_{y}$, only the doublet, belonging to $G/H$, has zero-mode, which we identify with our Higgs: $H = (A_{y}^{1(0)} + i A_{y}^{2(0)}, A_{y}^{4(0)} + i A_{y}^{5(0)})^{t}$. 
More explicitly the remaining zero-modes are 
\be
A_{\mu}^{a(0)} \ \ (a = 3,6,7,8); \ \ 
A_{y}^{a(0)} \ \ (a = 1,2,4,5);  \ \ 
\psi_{1R}^{(0)}, \ \ 
\psi_{2L}^{(0)}, \ \ 
\psi_{3L}^{(0)}.  
\ee  
It turns out that the remaining fields are just what we need in the standard model. Interestingly, the members of triplet fermion is known to have quantum numbers to be identified with quarks {\cite{KLY1}}. 

We now discuss the spontaneous gauge symmetry breaking and dynamical fermion mass generation due to the VEV of $A_{y}$, i.e. the Hosotani mechanism. The procedure itself is the same as in the case of SU(2), though the calculation becomes a little more complicated. So we just outline the results below. 
We first note that,  utilizing the global SU(2) $\otimes$ U(1) symmetry, we can always set the VEV in the form of $B_y^a=(B_y^1, 0, 0, 0, 0, 0, 0, 0)$.

Following the procedure explained in the SU(2) model, including the zero-mode contributions properly, we obtain the contribution from the gauge sector, i.e. 4D vector, scalar and ghost fields: 
\bea 
V_{eff}^{g+gh+s} &&= \frac{1}{2 \pi R} \int \frac{d^4 p_E}{(2 \pi)^4} \times \half \cdot 3 \nonumber  \\ 
                 && \cdot {\displaystyle{\sum_{n=-\infty}^{\infty}}}
[\log(\pe2+\NR)+\log(\pe2+(\frac{n-\alpha}{R})^2)+
2\log(\pe2+(\frac{n-\alpha/2}{R})^2)].
\eea  
Then the finite $\alpha$-independent part is calculated to be 
\be
V_{eff}^{g+gh+s}=-3C{\displaystyle{\sum_{n=1}^{\infty}}} \frac{1}{n^5}
(\cos (2\pi n \alpha)+ 2\cos( \pi n \alpha)).
\ee 
The contribution from the triplet fermions can be regarded as the sum of the contributions from $\psi_{1}$ and $\psi_{2}$ under the background $B^{1}_{y}$, which is just the same as in the SU(2) model,  and $\psi_{3}$ without the influence of $B^{1}_{y}$. Thus the contribution from the fermion system can be rather trivially evaluated: 
\be 
V_{eff}^f
= - \mbox{N}_{f} \cdot \frac{1}{2 \pi R} \int \frac{d^4 p_E}{(2 \pi)^4}
\half \cdot 4 [ {\displaystyle{\sum_{n=-\infty}^{\infty}}} 
\log[(\pe2+(\frac{n-\alpha/2}{R})^2+\half 
\log(\pe2+\frac{n^2}{R^2}) ].
\ee 
The finite $\alpha$-independent part is calculated to be the same as in the case of SU(2) (shown in Fig.1-(b)): 
\be
V_{eff}^f=4 \mbox{N}_{f} C{\displaystyle{\sum_{n=1}^{\infty}}} \frac{1}{n^5}
 \cos( \pi n \alpha).
\ee

\begin{figure}
\centerline{
\epsfxsize=6.5cm
\epsfbox{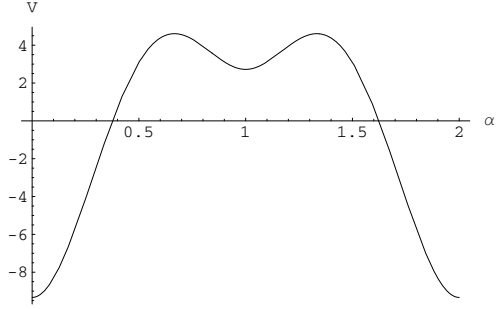}}
\caption{The contribution to the effective potential from the gauge-ghost system  in the SU(3) model, in the unit of $C=\frac{3}{128 \pi^7 R^5}$ as a function of $\alpha = g B^{1}_{y} R$.}
\end{figure}

The sum of all contributions to the effective potential now reads as 
\be
V_{eff}=-3C{\displaystyle{\sum_{n=1}^{\infty}}} \frac{1}{n^5}
[\cos (2\pi n \alpha)+ 2\cos( \pi n \alpha)]
+4{\rm{N}_f}C{\displaystyle{\sum_{n=1}^{\infty}}} \frac{1}{n^5}
\cos( \pi n \alpha) .
\ee

Now having the effective potential, we are ready to discuss the VEV and the Hosotani mechanism. It is interesting to note that the pattern of gauge symmetry breaking crucially depends on the number N$_{f}$ of triplet fermions, in
contrast to the case of the SU(2) model. 
For pure Yang-Mills theory, we see from Fig.2 that the vacuum is
located only at $\alpha=0$.  However, as ${\rm{N}_f}$ increases, as Fig.1-(b) suggests, the local minimum at $\alpha = 1$ becomes deeper. Thus for sufficient number of fermions we expect non-vanishing VEV. 
In fact, the difference of the depths at two local minima is given by  
\be
V_{eff}(\alpha=0)-V_{eff}(\alpha=1)
=4C{\displaystyle{\sum_{n=1}^{\infty}}} \frac{1}{(2n+1)^5}
(2{\rm{N}}_f-3).
\ee
Therefore, if ${\rm{N}_f}$ is less than two,
the global minimum is at $\alpha=0$ and the Hosotani mechanism does not work, though we get a ``Higgs" mass-squared 
\be
m^2_{A_y}=\frac{3g_{4}^2}{32 \pi^4 R^2}(9-{\rm{N}}_f)\zeta(3). 
\ee
On the other hand, for ${\rm{N}}_f \geq 2$ the global minimum is at $\alpha=1$,  i.e. $B_{y}^{1} = 1/(g R)$. In contrast to the case of SU(2) model, this VEV 
does break gauge symmetry spontaneously, but not into U(1): 
\be 
\mbox{SU(2)} \otimes \mbox{U(1)} \rightarrow \mbox{U(1)} \otimes \mbox{U(1)}.  
\ee
This is because for $\alpha = 1$ the Wilson loop 
\be 
W = 
\exp(ig \int_{-\pi R}^{\pi R} B_{y}^{1} T^{1}) 
=\pmatrix{-1 & 0 & 0 \cr
          0 & -1 & 0 \cr
          0 & 0 & 1
          } 
\ee
does not commute with $T^{6}, T^{7}$, but still commutes with $T^{3}$ and $T^{8}$. This VEV also causes a dynamical fermion mass. It is easy to understand that the VEV $B_{y}^{1} = 1/(g R)$ provides a Dirac mass term for the zero modes $\psi_{1R}^{(0)}$ and  $\psi_{2L}^{(0)}$, just as in the case of SU(2) model,   
\be 
ig \bar{\psi^{(0)}} \gamma_5 B^{1}_{y} T^{1} \psi^{(0)} = 
-i \frac{g}{2} B^{1}_{y}[\bar{\psi_{1R}^{(0)}} \psi_{2L}^{(0)} - \bar{\psi_{2L}^{(0)}} \psi_{1R}^{(0)}],  
\ee
which gives a massive Dirac particle with dynamical mass $1/(2R)$, while $\psi_{3L}^{(0)}$ remains massless. 
In this case the Higgs mass-squared is given as  
\be
m^2_{A_y}=\frac{9g^{2}_{4}}{128 \pi^4 R^2}(5+2{\rm{N}}_f)\zeta(3). 
\ee 
We summarize these results in Table 1.

\vspace{.5cm}

\begin{table}[h]
\begin{center}
\begin{tabular}[h]{|c|c|c|}
\hline
 & pattern of gauge symmetry breaking & dynamical fermion mass \\
\hline
 $(N_f \le 1)$ &
SU(2) $\otimes$ U(1) $\rightarrow$ SU(2) $\otimes$ U(1) &
0 \\
$(N_f \ge 2)$ & SU(2) $\otimes$ U(1) $\rightarrow$ U(1) $\otimes$ U(1) &
$\frac{1}{2R}$ \\
\hline
\end{tabular}
\caption{The pattern of gauge symmetry breaking and dynamical fermion mass}
\end{center}
\end{table}

\section{Conclusions and Remarks}

In this paper we studied the Hosotani mechanism, i.e. spontaneous gauge symmetry  breaking and dynamical mass generation of bulk fermions, in the bulk gauge 
models SU(2) and SU(3) on the space-time $\mbox{M}^{4} \times (\mbox{S}^1/\mbox{Z}_2)$. In addition to the gauge fields, matter fermions belonging to the fundamental repr.  were included. In these models extra-space component of a gauge field $A_{y}$ is identified with a 4D scalar field, ``Higgs", i.e. gauge Higgs unification has been realized. The crucial ingredient is the non-trivial assignment of Z$_2$-parity for the fields, forming an irreducible repr. of gauge group {\cite{Kawamura, Kawamura2}}. The parity assignment, ``orbifolding",  explicitly breaks the gauge symmetry and leads to a specific selection rule of the zero-mode (massless) fields in a given repr. at the classical level. 
Then some of $A_{y}$, associated with broken generators,  are assigned even parity and allowed to have zero-modes, which we identify with the Higgs, and generally develop non-vanishing VEV. 
The VEV $\langle A_{y} \rangle$ can be dynamically determined by the minimization of the radiatively induced effective potential. Such obtained VEV causes spontaneous breaking of gauge symmetry {\cite{Hosotani}}, which is left after the orbifolding, and/or generates dynamical masses for the zero-mode fermions (and Higgs itself \cite{HIL}). 

As the prototype model, we first discussed SU(2) model, and calculated the effective potential, by use of background field method, in order to fix the VEV. The generated VEV gave a Wilson loop $W = - I$, proportional to the unit matrix, thus leading to no spontaneous gauge symmetry breaking, while it gave a dynamical mass for doublet zero-mode fermion of the order $1/R$ ($R$: the radius of S$^{1}$),  as $W \neq I$. Such dynamical mass generation was argued to occur very similarly for the matter scalars belonging to the fundamental repr.. Thus the doublet-triplet splitting realized by the non-trivial Z$_2$-parity assignment in the SU(5) GUT {\cite{Kawamura}} is expected to be spoilt by the induced huge dynamical mass for the doublet scalar of the order $1/R \sim M_{GUT}$. 

As the candidate to provide a realistic model to incorporate the standard model  SU(2) $\times$ U(1), we next discussed SU(3) model with a non-trivial Z$_2$-parity assignment $P = \mbox{diag} (1, -1, -1)$. The orbifolding caused an explicit gauge symmetry breaking 
SU(3) $\rightarrow$ SU(2) $\times$ U(1), together with a specific selection rule of the zero-mode fermions. The resultant zero-modes of 4D vector, scalar and fermion fields are just what we need in the standard model, though some of matter fermion fields are still missing. In particular, we get a doublet 4D scalar, which is nothing but our Higgs field. We have shown that for sufficient number of matter fermions, the non-vanishing VEV of the Higgs doublet is radiatively generated. The induced VEV was argued to cause a  spontaneous breaking, SU(2) $\times$ U(1) $\rightarrow$ U(1) $\times$ U(1) and  dynamical Dirac mass for a pair of zero-mode Weyl fermions, in complete similarlity with ordinary mass generation via Yukawa coupling in the standard model. 
The derived model has very characteristic matter contents and various properties, as we will discuss in our forthcoming paper for the detail {\cite{KLY1}}. In particular the triplet matter fermions turn out to have the same quantum numbers as those of quarks, $(d_{R}, u_{L}, d_{L})^{t}$. This is in clear contrast to 
the ordinary 4D SU(3) unified gauge theory. In the 4D theory, 
the fermion triplet can be identified with $(e^{+}, \nu_{e}, e^{-})^{t}_{L}$. 
Also the Higgs, providing the mass for the electron belongs to the repr. 
$3 \times 3 = \bar{3} + 6$, but not 8 of SU(3). The essential difference between our 5D SU(3) model and the 4D SU(3) model is the difference of the chirality assignment for each component of the triplet fermion, which originates from our non-trivial assignment of the Z$_2$-parity in the orbifolding.  

The SU(3) theory, however,  still has some fundamental problems to be settled before it becomes a realistic model. First it should be pointed out that the Hosotani mechanism {\cite{Hosotani}} discussed in this paper did not lead to a reduction of the rank of the gauge group: U(1) $\times$ U(1) still has rank 2. In the forthcoming paper {\cite{KLY1}} we will demonstrate that to reduce the rank, 
necessary to realize U(1)$_{em}$, is possible once we introduce matter fields with adjoint repr.. We will also see, among other things, that suitably choosing the number of fermions belonging to fundamental and adjoint repr., we can realize a mild hierarchy between the mass scales generated by orbifolding and the Hosotani mechanism, which are, roughly speaking, both of the order $1/R$, as we 
discussed in the introduction. Such mild hierarchy is needed for the 2 step breaking, SU(3) $\rightarrow$ SU(2) $\times$ U(1) $\rightarrow$ U(1).

The bulk gauge theories, similar to the ones discussed in the present paper, 
have been already discussed by several authors in various context {\cite{ABQ, SUSY}}, where, however,  the dynamical fermion mass generation is not discussed 
and/or supersymmetry plays a crucial role.

%
%
%
\subsection*{Acknowledgment}

We would like to thank M. Sakamoto for useful comments and discussions.
The work of C.S.L. was supported in part by the Grant-in-Aid for Scientific Research of the Ministry of Education, Science and Culture, No.12047219, No.12640275.


\begin{thebibliography}{99}

\bibitem{Arkani}
N. Arkani-Hamed, S. Dimopoulos and G. Dvali, 
Phys. Lett. {\bf B429} (1998) 263; 
I. Antoniadis, N. Arkani-Hamed, S. Dimopoulos and G. Dvali, 
Phys. Lett. {\bf B436} (1998) 257. 
\bibitem{RS1}
L. Randall and R. Sundrum, 
Phys. Rev. Lett. {\bf 83} (1999) 3370;
Phys. Rev. Lett. {\bf 83} (1999) 4690; 
For a little different approach, see M. Kubo and C.S. Lim, a paper in preparation.
\bibitem{HIL}
H. Hatanaka, T. Inami and C. S. Lim, 
Mod. Phys.  Lett. {\bf A13} (1998) 2601. 
\bibitem{Daemi}
G. Dvali, S. Randjbar-Daemi and R, Tabbash, hep-ph/0102307. 
\bibitem{Georgi}
N. Arkani-Hamed, A.G. Cohen and H. Georgi, Phys. Lett. {\bf B513} (2001) 232. 
\bibitem{Nomura}
L. Hall, Y. Nomura and D. Smith, hep-ph/0107331. 
\bibitem{ABQ}
I. Antoniadis, K. Benakli and M. Quiros, hep-th/0108005.
\bibitem{Kawamura}
Y. Kawamura,
Prog. Theor. Phys. {\bf 103} (2000) 613.
\bibitem{Kawamura2}
Y. Kawamura, 
Prog. Theor. Phys. {\bf 205} (2001) 691;
Prog. Theor. Phys. {\bf 205} (2001) 999.
\bibitem{Rub-Sha}
K. Akama, ``Pregeometry" in Lecture Notes in Physics, 176, Gauge Theory and Gravitation, 
Proceedings, Nara, 1982, ed. by K, Kikkawa, N. Nakanishi and H. Nariai, (Springer-Verlag, 1983), 267-271;  
V. A. Rubakov and M. E. Shaposhinikov
Phys. Lett. {\bf B125} (1983) 136.
\bibitem{DS}
G. Dvali and M. Shifman,
Nucl. Phys. {\bf B504} (1997) 127.
\bibitem{Hosotani}
Y. Hosotani,
Phys. Lett. {\bf B126} (1983) 309;
Phys. Lett. {\bf B129} (1983) 193;
Ann. Phys.  {\bf 190}  (1989) 233.
\bibitem{Hatanaka}
H. Hatanaka,
Prog. Theor. Phys {\bf 102} (1999) 407.
\bibitem{KLY1}
M. Kubo, C. S. Lim and H. Yamashita,
in preparation.
\bibitem{SUSY}
R. Barbieri, L. J. Hall and Y. Nomura,Phys. Rev {\bf D63} (2001) 105007; hep-th/0107004; 
N. Arkani-Hamed, L. Hall Y. Nomura, D. Smith and N. Weiner, Nucl. Phys. {\bf B605} (2001) 81; 
L. J. Hall and Y. Nomura, Phys. Rev {\bf D64} (2001) 055003; 
Y. Nomura, D. Smith and N. Weiner, Nucl. Phys {\bf B613} (2001) 147.
%
%
%
\end{thebibliography}
\end{document}